# BEAM DIAGNOSTICS IN THE ADVANCED PLASMA WAKEFIELD EXPERIMENT AWAKE

A.-M. Bachmann[1,2]*, Max Planck Institute for Physics, Munich, Germany
P. Muggli, Max Planck Institute for Physics, Munich, Germany
[1]also at TU Munich, Munich, Germany
[2]also at CERN, Geneva, Switzerland
on behalf of the AWAKE Collaboration

*Abstract*

In AWAKE a self-modulated proton bunch drives wakefields in a plasma. Recent experiments successfully demonstrated many aspects of the self-modulation of the drive bunch as well as acceleration of test electrons. Next experiments will focus on producing a multi-GeV accelerated electron bunch with low emittance and low energy spread. The experiment requires a variety of advanced beam diagnostics to characterize the self-modulated proton bunch at the picosecond time scale. These include optical transition radiation and a streak camera for short and long time scale detailed imaging of self-modulation and hosing, coherent transition radiation for modulation frequency measurements in the 100–300 GHz frequency range and multiple fluorescent screens for core and halo measurements. An overview of these diagnostics will be given.

## INTRODUCTION

AWAKE operates at one of the lowest plasma densities of all currently available plasma-based accelerator experiments. The plasma electron density $n_{e0}$ determines the fastest time-scale $\tau$ of characteristics of the bunch modulation with the plasma electron angular frequency, $\omega_{pe} = \left(n_{e0}e^2/\epsilon_0 m_e\right)^{1/2}$, with $\tau = 2\pi/\omega_{pe}$, where constants have usual meaning: $e$, charge of the electron: $\epsilon_0$, vacuum permittivity; $m_e$, mass of the electron; $c$, the speed of light in vacuum. For $n_{e0} = 10^{14}$–$10^{15}$ cm$^{-3}$ this corresponds to $\tau = 3.5$–11.1 ps. The time resolution needed to directly observe the modulation is thus at the limit of currently commercially available streak cameras ($\approx 200$ fs). The density also determines the smallest spatial scale through the cold plasma collisionless skin depth $c/\omega_{pe}$. The main phenomena to resolve are the structure of the proton bunch that drives plasma wakefields and of the wakefields themselves. The incoming proton bunch is much longer than the wakefields' period $\tau$. Thus it experiences self-modulation (SM) as it travels through the plasma [1]. The SM process acts on the bunch through the periodic focusing and defocusing transverse wakefields along the bunch, generating a train of micro-bunches with periodicity $\sim \tau$ and shorter than $\tau$.

To study this fundamental beam-plasma interaction process in detail, we acquire time-resolved transverse images of the modulated bunch density distribution measuring the optical transition radiation with a streak camera [2, 3]. With multiple fluorescent screens along the beam path, we acquire time integrated images of the modulated bunch to complement the time-resolved images [4]. Additionally, with heterodyne receivers the frequency of the modulation can be measured independently [5]. We measure energy, energy spread and charge capture of an electron bunch, externally injected and accelerated in the plasma with an electron spectrometer [6, 7].

We describe the diagnostics that were used in the first round of experiments (2016–2018). We also briefly outline the diagnostic challenges for the upcoming experiments (starting 2021) that will focus on the accelerated bunch quality. As with all plasma-based accelerators beam diagnostics must measure small spatial and temporal scales, typically µm and 100 fs, respectively. In addition, diagnostic measurements must be integrated into compact spaces and measure simultaneously properties of very different beams, the high population, long duration and long $\beta$-function proton bunch and the low population, short duration and small spatial size electron bunch. Moreover, in these experiments, they must overlap in space and time with $\ll \tau$ and $c/\omega_{pe}$ accuracy at the plasma entrance, an overlap that also complicates diagnostics at the plasma exit.

## THE AWAKE EXPERIMENT

In the AWAKE experiment the transverse central 2 mm diameter of a 10 m long rubidium (Rb) vapor source is ionized by a 4 TW (peak power) Ti:Sapphire laser pulse. An $\approx 12$ cm-long CERN SPS proton bunch propagates through the plasma. The long proton bunch drives wakefields in the plasma. Focusing and defocusing transverse fields alternate along the bunch with the wakefields' period $\sim \tau$. The wake-fields act back on the long proton bunch, leading to radial modulation. The modulation is seeded with the ionizing laser-pulse co-propagating within the proton bunch (Seeded Self-Modulation) [8]. In the focusing phase of the wakefields protons remain on axis, while in between in the defocusing phase protons diverge, resulting in micro-bunching. The modulated proton bunch resonantly drives wakefields. The longitudinal wakefields accelerate a short electron bunch, generated on a photo-cathode and externally injected [9]. Beam diagnostics are placed downstream of the plasma to characterize the modulated proton bunch and the accelerated electrons. The fluorescent screens to measure the time integrated transverse distribution of the proton bunch are shown in Fig. 1, referred to as imaging station 1 and 2. The radial extent of defocused proton distribution depends on the transverse wakefields' amplitude they have experienced [4]. Two thin metal screens (OTR, CTR screen in Fig. 1) allow for the measurement of the optical transition radiation (OTR) and coherent transition radiation (CTR) emitted by the protons and micro-bunches, respectively, when entering the screens. The OTR is sent to a streak camera, acquiring a time-resolved image of the transverse distribution of a slice of the modulated proton bunch. The CTR is detected by heterodyne receivers to determine the modulation frequency of the bunch train. The charge and energy of the electron bunch after acceleration in the plasma are measured with the electron spectrometer.

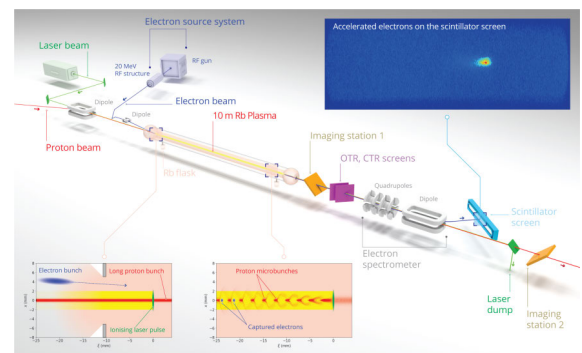

Figure 1: Layout of the AWAKE experiment [6].

---

* bachmann@mpp.mpg.de

## PROTON BUNCH DIAGNOSTICS

In the following we describe the diagnostics used to characterize the modulated proton bunch driving the plasma wakefields and we show some example results.

### Optical Transition Radiation and Streak Camera for Time-Resolved Transverse Distribution Measurements

The modulated bunch propagates through a foil (280 μm silicon wafer coated with 1 μm mirror-finished aluminium) where OTR, carrying the spatio-temporal pattern of the proton bunch distribution, is emitted [2]. The screen is placed 3.5 m after the plasma exit (see Fig. 1). The light is relay imaged onto the slit (20 μm) of a streak camera (Hamamatsu C10910-05) acquiring the time-resolved transverse bunch density distribution. These images can display the long time evolution of the self-modulated train and defocused protons. The evolution of the protons leaving the bunch is captured only over a few wakefields' periods due to aperture limitations of the imaging system (field of view $x \approx \pm 4$ mm in Fig. 2).

To reach the best time resolution, the camera must be operated in a speckle regime. The time resolution then reaches $\approx 0.8$ ps. However, we image the transverse bunch density distribution and operate with larger intensity images in order to obtain single images of the bunch structure. The OTR signal (equal in duration to that of the proton bunch) lasts for $\approx 1$ ns, much longer than the shortest time window (73 ps) with maximum time resolution, potentially further decreasing the time resolution. This typically decreased the time resolution to $\approx 2$ ps for the short time window.

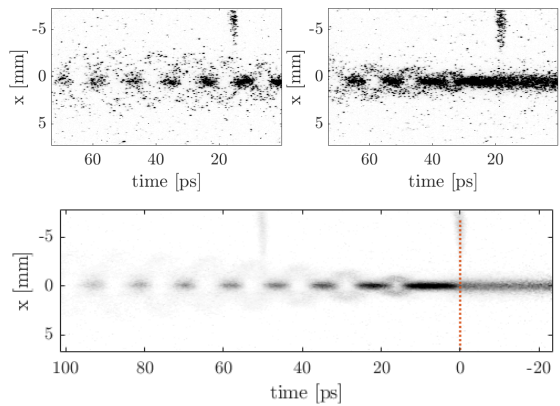

Figure 2: Single streak camera image of the start of the SM (top right), single streak camera image with a 50 ps time delay to the start of the SM (top left) and sets of the two time delays of 20 streak camera images each, stitched together using the timing reference signal (bottom).

To decrease the noise of the images, we average several acquisitions of the same position along the bunch from successive events. This requires a repetitive signal (reproducibility of the modulation) and a very stable trigger for the diagnostic ($< \tau$) or an independent signal tied in time to the modulation itself. The electronic trigger jitters on the tens of picoseconds scale. We therefore send a mirror bleed replica of the laser pulse that ionizes the Rb vapor, and thus starts the self-modulation of the bunch, to the streak camera together with the OTR. We offset it from the OTR signal along the camera slit (see signal at $x \leq -4$ mm in Fig. 2). This time reference signal has an accuracy of 0.6 ps with respect to the ionizing laser pulse and thus with the start of the self-modulation. This replica combined with a tunable delay, serves as a time reference that can be used to average and stitch multiple events [10]. The averaging of multiple events allows for images of the modulated bunch with a much higher signal to noise ratio than with single images. By stitching the images of the shortest time window, i.e. combining a number of image sets taken with various reference signal delays with respect to the start of the self-modulation, we can measure fast time scale details along the bunch, as well as examine modulation structures with greater time resolution over a long time scale (see Fig. 2).

At the location of the OTR screen, the RMS transverse size of the proton bunch is $\approx 580$ μm and the transverse size of the micro-bunches is similar [3]. Measurements show that the spatial resolution is on the order of 180 μm.

Figure 2 shows an example of a single image of the proton bunch near the start of the SM (top right), a single image with camera trigger and reference laser pulse shifted by 50 ps (top left) and of two sets of images (20 events each), acquired with a 50 ps trigger delay in between and "stitched" together (bottom). Here the first reference signal timing is without delay with respect to the ionizing laser pulse, marked with the red dashed vertical line in the stitched image. This corresponds to the start of the plasma, here at time $t = 0$. In this case the plasma electron density is $n_{e0} = 0.9 \times 10^{14}$ cm$^{-3}$ and thus $\tau = 11.7$ ps.

We use the space and time-resolved transverse slice of the modulated proton bunch density mapped from the streak camera images for qualitative and quantitative characterization of the self-modulated bunch. The position, shape and length of the proton micro-bunches is crucial for plasma wakefield acceleration experiments. The driver charge (here protons) should ideally be in the decelerating and focusing phase of the wakefields (one fourth of the plasma period in linear wakefield theory [11]). This would lead to large plasma wakefields' amplitudes and thus large acceleration and focusing of injected particles when placed at the right longitudinal position within the wakefields' period.

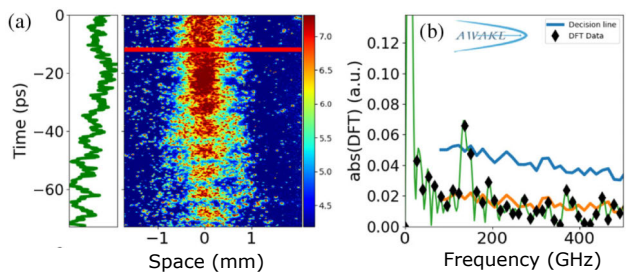

Figure 3: Image a) shows a streak camera image of the self-modulated proton bunch (moving from the bottom to the top) with the plasma starting at the red horizontal line and the projection of the central density distribution with the green curve on the left, b) shows the DFT of the projection, with determined peaks marked with black diamonds and the dominant frequency above the noise threshold determined as 137 GHz with $n_{e0} = 2.5 \times 10^{14}$ cm$^{-3}$ (from [3]).

With the time-resolved bunch density maps the modulation frequency can be determined using the discrete Fourier transform (DFT) of the modulated bunch train [3]. This is shown in Fig. 3, with the streak camera image of a single event and the projection of the central bunch density in green on the left and the DFT and peak determination on the right. With a plasma period of $\tau < 11$ ps and a resolution of $\approx 2$ ps, an upper bound of the micro-bunch length can be determined. The relative charge per micro-bunch can be measured, taking into account the effect of the streak camera slit on the transverse distribution [12]. This is particularly important for example when studying the effect of plasma density gradients on the SM process and validate theory and simulation predictions [13]. Competing instabilities of a particle bunch in the plasma, as the hosing instability, can be analyzed by studying the non-axi-symmetric behaviour of the modulated bunch with the time-resolved images [14]. In all studies the 3.5 m of vacuum propagation between the plasma exit and the OTR screen, and thus additional divergence of protons, needs to be taken into account.

*Fluorescent Screens for Time Integrated Transverse Distribution Measurements*

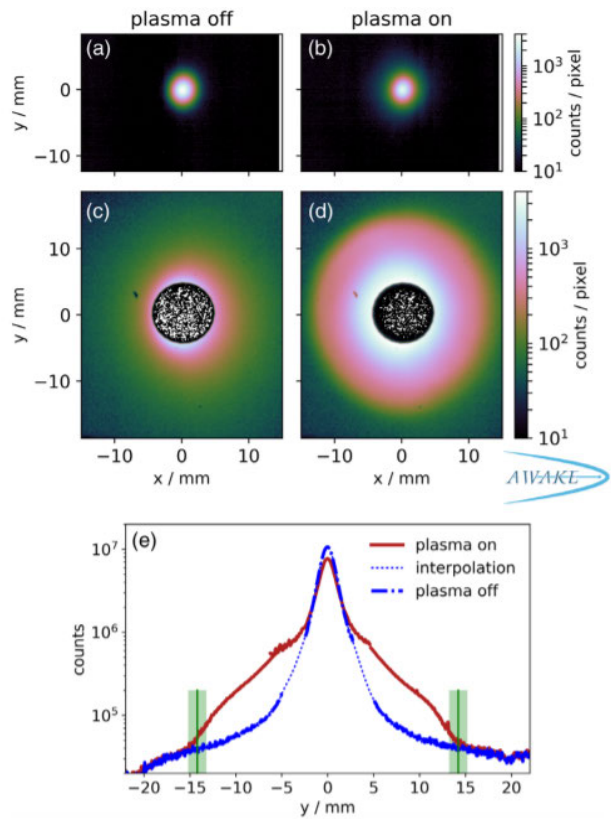

Figure 4: Time integrated transverse proton bunch density distribution acquired with fluorescent screens for the measurement of defocused protons and transverse wakefields' amplitude determination (from [4]). Images a), b) show core images, c), d) halo images for the same events with a), c) plasma OFF and b), d), plasma ON. Image e) shows the reconstructions of the bunch profile from the two cameras, as well as the value determined for the halo radius (green lines) and its uncertainty (green shaded regions). These images were obtained at imaging station 2 ($\approx 10$ m from the plasma exit). Similar images were obtained at imaging station 1.

Fluorescent screens are placed along the beam line before and after the plasma, measuring the time integrated transverse proton bunch distribution, see Fig. 4. The radial modulation of the bunch in plasma leads to divergence of protons in defocusing regions between the focused regions of the micro-bunches. These protons appear as a halo around the bunch core on the imaging stations $\approx 2$ m and $\approx 10$ m after the plasma. The halo's radial extent is a measure of the integrated defocusing wakefields the protons have experienced [4]. At each of the two imaging stations, the light is divided into two light paths with a camera each. One camera measures the entire bunch distribution with parameters adjusted to the high density core of the bunch. This image also gives information about the beam alignment and eventual kicks of the beam by the plasma when beam and plasma are misaligned. For the second camera the light passes through a mask, blocking the high intensity light from the dense core. Parameters are adjusted to measure the low light level halo. The combination of the two images enables a larger dynamic range measurement of the entire radial bunch distribution, detecting for the same event the significantly lower intensity of the signal from defocused protons (halo) around the core without saturation. The halo measurements have successfully shown that wakefields grow along the proton bunch [4], as predicted by theory and simulation results. SM, an axi-symmetric bunch-plasma mode of interaction produces a circular core and halo, when the finite radius plasma and the proton beam are well aligned. On the contrary, the hosing instability [15] is a non-axi-symmetric mode of interaction that may produce correspondingly asymmetric core and halo. Some of the self-modulation and hosing characteristics can thus be studied from time-integrated and time-resolved images of the proton bunch.

*Heterodyne Detectors for Modulation Frequency Measurements Using Coherent Transition Radiation*

We obtain an independent frequency measurement of the proton bunch modulation by analysing the CTR emitted when the bunch train enters a foil (Al-coated silicon oxide screen) placed after the plasma (see Fig. 1). The CTR signal is in the $f_{CTR} \approx 100-300$ GHz frequency range for plasma electron densities in the $n_{e0} \approx 10^{14}$–$10^{15}$ cm$^{-3}$ range. This microwave signal is transported with waveguides to heterodyne detectors. They consist of a frequency-tunable, local oscillator synthesizer, an amplifier/frequency multiplier chain, a sub-harmonic mixer and an oscilloscope [5]. The signal at $f_{CTR}$ is mixed with a reference signal at approximately the same frequency $f_{ref} \cong f_{CTR}$, guessed from the rubidium, thus plasma density $f_{CTR} \cong f_{pe} \sim n_{e0}^{1/2}$. The difference frequency signal at $f_{IF} = |f_{CTR} - f_{ref}|$ is acquired with a fast, multi-GHz bandwidth oscilloscope. The reference signal is generated by the frequency multiplication of a tunable local low frequency oscillator ($f_{LO}$) as $f_{ref} = n_{harm} f_{LO}$, with harmonic number $n_{harm}$. One can then determine $f_{CTR} = n_{harm} f_{LO} \pm f_{IF}$ from measurements of $f_{IF}$ with multiple $f_{LO}$ with small frequency increments, for a fixed $f_{CTR}$.

In the experiment, the CTR signal is split in three and analyzed with three heterodyne systems covering the entire 100–300 GHz frequency range. For low plasma densities, this diagnostic provides simultaneous information about two or three of the modulation harmonics. Detection of harmonics of the modulation signal indicates deep modulation of the bunch radius and thus formation of a micro-bunch train with $\tau = 1/f_{CTR}$ period. The heterodyne systems provide short (~duration of the proton bunch), low frequency (5–20 GHz) signals on the oscilloscope, consisting of only a few periods. We use DFT to determine $f_{IF}$.

The modulation frequency of the proton bunch has been measured with this technique and compared to the results obtained from DFT analysis of streak camera images and the plasma frequency expected from the plasma density (when measuring the vapor density and assuming full ionization of the Rb valence electron) [3]. It was also used to measure the modulation frequency as a function of plasma density gradients [13]. There is very good agreement between the modulation frequencies obtained from DFT of streak camera images time profiles and from CTR signals. Moreover, CTR signals at the second harmonic of the modulation frequency were also detected, confirming the deep modulation of the bunch density.

## ELECTRON BUNCH DIAGNOSTICS

The electron bunch is externally injected into the wakefields, driven by the proton bunch. We characterize the accelerated electron bunch after the plasma with the spectrometer.

*Spectrometer for Charge, Energy and Energy Spread Measurements*

For electron injection, temporal synchronization between the self-modulation of the driver and the short electron bunch is ensured by emitting the electrons from a photo-cathode with a frequency-tripled fraction of the ionizing laser pulse. They are accelerated from the RF-cathode through a booster structure and injected into the plasma (oblique or on axis injection) with initially $\approx 18$ MeV and accelerated up to multiple GeV [6]. At injection, the bunch charge is $\lesssim 1$ nC and the RMS length estimated to be $\gtrsim 5$ ps [16], thus on the order of the wakefields' period. This injection scheme thus corresponds to injecting electrons at many phases of the wakefields ($2\pi$) and letting the wakefields capture and accelerate a fraction of the incoming bunch population. The electrons may dephase through energy gain and loss.

In the current setup, after the plasma exit the electrons are focused with two quadrupoles and bent with a dipole onto a scintillator screen (see Fig. 1), imaged onto a camera in a dark room [17].

The dipole introduces a correlation between the horizontal position and energy of the electrons on the screen. Thus the peak energy and the energy spread of the accelerated electron distribution can be measured [6, 16].

The relation between light and charge was acquired exposing the screen directly with a beam with known charge [18]. Using the light-charge calibration one can determine the charge after the plasma, and thus capture rate of the injected electrons [6].

## DIAGNOSTICS FOR FUTURE EXPERIMENTS

In previous experiments, the goal was to demonstrate and study the self-modulation of the proton bunch in plasma and to show that electrons can be accelerated in wakefields driven by the resulting bunch train. The electron bunch length was comparable to the wakefields' period to ensure covering all phases of the wakefields, ensuring that electrons would be captured for all events, regardless of the relative timing of electrons with respect to the wakefields. In addition, electrons were injected at an angle with respect to the proton bunch propagation axis.

In upcoming measurements [19], the focus will be on preserving the quality of the accelerated electron bunch, such as high charge, low emittance and low energy spread. Those can only be achieved with a short electron bunch ($\ll \tau$) injected at the right phase of the wakefield, focusing and accelerating. The electron bunch parameters must be tailored such that plasma electron blow-out can be reached with a transverse size matched to the ion column focusing force to preserve slice emittance. Beam loading needs to be reached, to obtain a narrow energy spread and preserve the projected emittance [20].

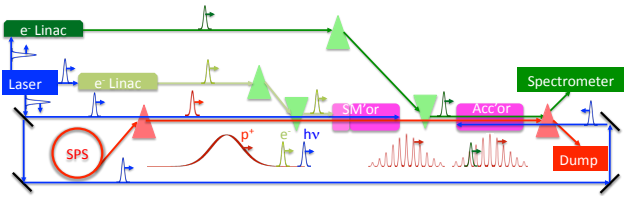

Figure 5: AWAKE setup planned for future measurements.

During the SM growth the phase of the wakefields evolves [21, 22], i.e. their phase velocity is slower than the proton bunch velocity [23]. Electrons will thus be injected after the proton bunch is fully modulated. Therefore two plasma stages, separated by a short gap for electron injection will be used, as shown in Fig. 5. The first plasma stage is for the SM of the long proton bunch, the second plasma stage for the acceleration of the electron bunch in the wakefields driven by the fully modulated proton bunch. The first plasma stage will include a plasma density step for wakefields to maintain amplitudes comparable to that at their saturation value, over long distance [13]. We want to monitor the effect of the density step on the plasma electron density perturbation sustaining wakefields at the end of the plasma with a new THz shadowgraphy diagnostic.

These experiments will bring new diagnostics challenges, both on the entrance and exit side of the plasma. The matched $\beta$-function of the $\approx 150$ MeV electron beam to an ion column of density $10^{14}$–$10^{15}$ cm$^{-3}$ is $\beta_{0m} \approx 4$–13 mm. With a normalized emittance of 2 mm·mrad this corresponds to a waist transverse size of $\sigma_r \approx 5$–10 µm, very challenging to produce and measure, especially in the Rb vapor environment. The bunch length $\sigma_\xi$ on the order of 60–100 µm can be measured by electro-optic sampling or CTR interferometry. It can be monitored on a single-event-basis by measuring the CTR energy $E_{CTR}$ and the bunch population $N$, and using the dependence $E_{CTR} \sim N^2/\sigma_\xi$. The spatial alignment between the two beams will be measured and monitored in the presence of the two bunches with very different parameters (population, $\beta$-function, size and length) using conventional (protons) and new Cherenkov diffraction (electrons) beam position monitors [24].

At the plasma exit, the bunch energy spectrum will be measured using a standard imaging magnetic spectrometer (quadrupoles, dipole). The accelerated electron and the modulated proton bunch overlap in space and time. Thus measuring the electron bunch emittance requires a diagnostic that separates the two bunches or the two signals. In the first category, measuring the transverse size in the non-dispersive plane of the magnetic spectrometer, the equivalent of a quadrupole scan in a single or multiple events, can in principle yield emittance [25]. In the second category, measuring the betatron radiation by the accelerated electron bunch can yield an emittance measurement [26]. Also diffraction radiation measurements, taking advantage of the difference in relativistic factor between the accelerated electrons and the protons, may provide an emittance and other bunch measurements [27, 28].

## CONCLUSION

AWAKE is the first proton-beam-driven wakefield accelerator [8]. It proved the concept of seeded self-modulation and the acceleration of externally injected electrons up to 2 GeV over 10 m of plasma.

The experiment operates with a low plasma density, putting the typical time scale of observation in the few picoseconds regime. We use OTR and a streak camera with a reference laser signal to obtain detailed images of the proton bunch self-modulation over hundreds of picoseconds while preserving picosecond resolution. Stitching many events together greatly suppresses the noise of the streak camera images. We also use heterodyne detection of CTR emitted by the bunch train to directly measure the modulation frequency, including its harmonics.

Together with more standard beam diagnostics, including screens and energy spectrometer, these diagnostics have allowed for a detailed understanding of the SM process [3, 4, 13, 16] and for the demonstration of acceleration of externally injected electrons to GeV energies [6].

Future experiments will focus on producing an accelerated electron bunch with significant charge, low relative energy spread, while essentially preserving its incoming emittance. These experiments present a number of new challenges for diagnostics, which include: size and alignment measurements at the micron scale; synchronization at the sub-picosecond scale; single-event measurements of all incoming and outgoing parameters, including the emittance of the accelerated bunch.

Because of its typical operating frequency range (100–300 GHz), this accelerator pushes diagnostics typical of RF accelerators, operating in the GHz frequency range, towards those, typical of other plasma-based accelerators, operating at higher plasma densities and thus in the THz frequency range.